# Development of Carbon-Fiber Electrodes Modified for Electrochemical ATP Detection

## Karbon-Fiber Elektrotların Elektrokimyasal ATP Tespiti İçin Modifiye Edilerek Geliştirilmesi

Mustafa ŞEN

Biyomedikal Mühendisliği, İzmir Kâtip Çelebi Üniversitesi, Çiğli, İzmir, Türkiye, mustafa.sen@ikc.edu.tr

## Özet

Adenozin trifosfat (ATP) hücresel metabolizmanın devamlılığı için gerekli kimyasal enerjinin transferinde koenzim olarak kullanılan yüksek enerjili bir moleküldür. Ayrıca, hücre dışı ATP bağışıklık, sinir ve vasküler sistemlerde, hücreler arası iletişimde önemli bir göreve sahiptir. ATP daha çok ticari olarak ta elde edilebilen lusiferin-lusiferaz kitleri kullanılarak optik algılama temelli ölçülmekte ve bu metot doğası gereği lokal ölçüme olanak vermemektedir. Bu çalışmada temel olarak sahip oldukları küçük boyuttan dolayı hücre ve dokularda elektrokimyasal lokal algılama yapabilen karbon-fiber mikroelektrotlarının üretimi ve sonrasında elektrot yüzeylerinin gerekli enzimlerle modifiyesi şeklinde elektro aktif olmayan ATP molekülünün algılaması gerçekleştirilmiştir.

**Anahtar kelimeler:** Elektrokimyasal algılama, ATP, biyosensör, $H_2O_2$

## Abstract

Adenosine triphosphate (ATP) is a highly energetic molecule used as a coenzyme in the transfer of chemical energy required for the continuation of cellular metabolism. In addition, extracellular ATP has an important role in cell-to-cell communication in immune, nervous and vascular systems. ATP is detected mainly using commercially available luciferin-luciferase kits and this method does not allow local measurement. In this study, first carbon-fiber microelectrodes capable of local detection were fabricated and then, the surface of these electrodes were modified with enzymes for detection of non-electroactive ATP molecules.

## 1. Giriş

Klasik olarak adenozin trifosfat (ATP), hücre metabolizması için enerji kaynağı olarak görülür ve bu çerçevede hücre içi bazı enzimlerin aktivitesi için gerekli olan enerjiyi sahip olduğu kimyasal bağlar içinde barındırır [1]. Biyolojik para birimi olarak kabul edilen ATP basit ve kompleks karbonhidrat ve yağlardan hücre içi 3 değişik metabolik yolla üretilir; glikoliz, oksijenli solunum ve beta-oksidasyonu. Hücre bu molekülün yüksek enerjisini canlılığın devamı için birçok hücresel faaliyette kullanabilir; örneğin, kofaktör olarak bazı moleküllerin hücre içi aktif transportunda, kas kasılması esnasında aktin-miyozin çapraz köprü oluşma döngüsünde, protein kinazların post-translasyon modifikasyonda proteinlere fosfat aktarımında kullanımı gibi [2]. ATP'nin hücre içi rolü çok iyi karakterize edilmiş olup bu rolüne ek olarak hücre dışında sinyal molekülü olarak görev aldığı yaklaşık 50 yıl önce keşfedilmiştir [3]. Bu konuda yapılan çalışmalar ATP'nin birçok farklı hücre tarafından salındığı ve işlevini otokrin ve/veya parakrin sistemi aracılığıyla yürüttüğü anlaşılmıştır. Genelde ATP'nin deteksiyonu lusiferin-lusiferaz assay kullanılarak ölçülmekte ve bu metot doğası gereği lokal ölçüme olanak vermemektedir [4, 5]. Ek olarak diğer bazı dezavantajları da mevcuttur; örneğin, analiz edilen örnekten gelen lüminesens ölçülmek istenen ATP biyolüminesensi etkileyerek ölçüm kalitesini düşürebilmekte ve ayrıca, hassasiyet büyük oranda güçlendirilmiş kameralara ve bunların sağlayacakları çözünürlüğe bağlı olması bu tip sistemleri daha masraflı yapmaktadır. Kullanılan kimyasalların masraflı ve stabilitelerinin iyi olmadığı da unutulmamalıdır. Bahsi geçen dezavantajlar düşünüldüğünde elektrokimyasal algılama alternatif bir metot olarak göze çarpmaktadır. Şu ana kadar birçok elektrokimyasal temelli ATP deteksiyon sistemleri geliştirilmiş olup genelde deteksiyon için çeşitli enzimler ve aptamerler kullanılarak istenilen hassasiyete ulaşılmaya çalışılmıştır. Aptamer temelli sensörlerin çoğu iyi sonuç vermekle beraber bu sensörlerin bazı handikapları mevcuttur; örneğin, aptamer immobilizasyonu ve mikro çevrenin aptamer yapısına ve aptamer-ligand interaksiyonuna etkisi tam olarak bilinmemekte, aptamerlerin tuzlara karşı hassas oluşu nedeniyle analiz örneklerinin tuz seviyesi aptasensörlerin hassasiyetini etkileyebilmektedir. İlaveten bazı proteinler DNA aptamerlerine spesifik olmayan bir biçimde bağlanması veya örnekte bulunan nükleik asitlerin aptamerle hibridizasyon oluşturması aptamer konformasyonunu ve bağlanma bölgesini etkileyerek analit ölçüm hassasiyetini değiştirebilmektedir [6]. Ayrıca aptamerler genelde pahalı ve ilişkili teknolojinin yaygın olmaması bu tip biyosensörlerin pratik kullanımı için çeşitli problemler oluşturmaktadır [7]. Bu çalışmada elektrot yüzeyinin çeşitli enzimlerle modifikasyonuyla ATP algılaması gerçekleştirilmiştir. İlk olarak, karbon-fiber elektrotlar üretilmiş ve elektrot yüzeyleri

Bakır tellere bağlanan karbon fiberler cam kılcal tüplerine sabitlendikleri bakır teller yardımıyla tersten yerleştirilmiş ve karbon fiberlerin herhangi bir zarar görmemesi için bakır teller cam tüplerin üzerine ısı ile daralan makaron ile sabitlenmiştir. Cam kılcal tüpler

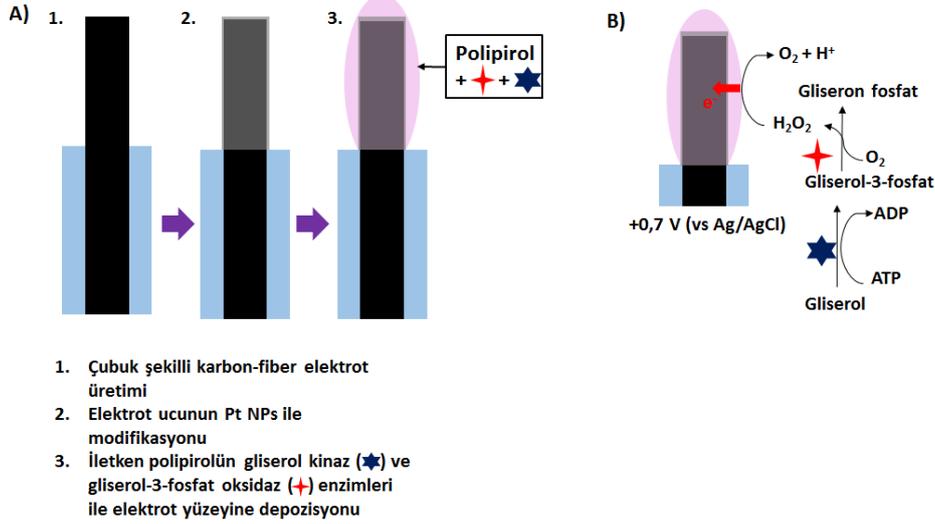

Şekil 1. İğne uçlu karbon-fiber elektrotlardan ATP biyosensörü üretimi (A) ve ATP algılamasında kullanımı (B)

elektrokimyasal olarak Pt nanopartikülleri ile modifiye edilmiştir. Sonrasında, elektrot yüzeyleri polipirol aracılığıyla çeşitli enzimler ile modifiye edilmiştir (Şekil 1A). Modifikasyonda kullanılan enzimlerin ATP yıkım yan ürünü olarak ortama saldıkları $H_2O_2$'nin elektro-oksidasyonuyla ATP miktar tayini yapılmıştır (Şekil 1B).

## 2. Yöntem ve Sonuçlar

KFE'ler çeşitli şekillerde üretilmekle birlikte bu çalışmada cam kılcallarının karbon fiberlere geçirilmesi şeklinde üretimi tercih edilmiştir. İlk olarak karbon fiberlerin cam kılcal tüplere geçirilmesi ve daha sonrasında karbon fiber elektrotların elektrokimyasal sisteme (potentiyostat) bağlanmasında kolaylık sağlaması gibi nedenlerden dolayı karbon fiberler sadece uç kısımları açılmış kaplı bakır tellere karbon fiberlere mikçekme makinesi (PC-10 puller, Narishige, Japonya) geçirilerek Pt-KFE'ler üretilmiştir; kullanılan çekme parametreleri: seçenek 1 ve 65 °C [8-11]. Elektrot ucunda çekme işlemi sonrası sarkan karbon-fiber tel bir jilet yardımıyla hassas bir şekilde kesilmiştir. Dışarıda bırakılan karbon-fiber tel boyutu elektrotun yüzeyini belirleyeceğinden aynı cevabı verebilen biyosensörler için bu kısmın olabildiğince benzer olması gerekmektedir. Bu noktada ilk olarak beyaz bir A4 kâğıdı üzerinde bir ölçek oluşturulmuş ve bu ölçek yardımıyla da elektrot yüzey alanı benzer olacak şekilde karbon-fiberler kesilmiştir. İşlemin başarıyla gerçekleşip gerçekleşmediği CV yardımıyla ortaya konulmuştur. İğne uçlu KFE'lerin CV eğrileri 1 mM FMA + PBS solüsyonunda elde edilmiş ve sonrasında eğriler karşılaştırılmıştır. CV işlemi için KFE'lerin potansiyeli 0 ile +0.5 V (vs. Ag/AgCl) arasında 50 mV/s hızında taranarak elde edilmiştir. Elde edilen CV eğrileri

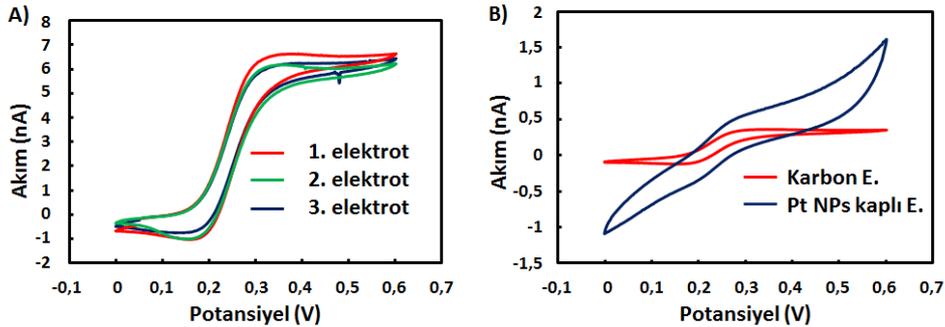

Şekil 2. Üç farklı elektrota ait FMA solüsyonunda elde edilen CV eğrileri (A) ve Pt modifikasyonu sonucunda akımda gözlenen farklılık (B).

bağlanmıştır. Bağlama işlemi için karbon fiberler ve bakır teller gümüş (Ag) pastası kullanılarak bir birlerine 180 °C derecede pişirilerek tutunması sağlanmıştır.

karşılaştırıldığında elektrotların olabildiğince benzer ve elektro aktif FMA'ya karşı da anlamlı yanıt verdiği gözlemlenmiştir (Şekil 2A). Bu sonuçlar kullanılan yeni stratejinin tekrar edilebilirliği yüksek iğne uçlu elektrotların üretiminde başarılı bir şekilde uygulanabileceğini kanıtlamaktadır. Bir sonraki işlemde elde edilen CV eğrisi modifiye edilmemiş KFE'nin eğrisi ile karşılaştırıldığında anlamlı bir farklılığın oluştuğu gözlemlenmiştir (Şekil 2B). Beklendiği üzere Pt

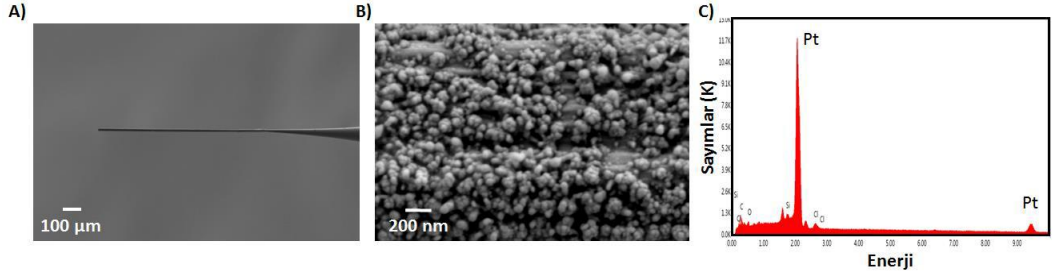

Şekil 3. Pt-KFE'ye ait uzak (A) ve yakın (B) SEM görüntüleri ve yakın görüntüde alan üzerine yapılan EDS analiz sonucu (C).

iğne uçlu elektrotların yüzeyi $H_2O_2$'nun elektrokimyasal ölçümü için Pt NP'ler ile modifiye edilmiştir [12]. Modifiye işleminde akım seviyesinin yüksek olması nedeniyle üçlü elektrot sistemi kullanılmıştır; (1) iğne uçlu KFE çalışma elektrotu, (2) karşıt elektrot (kalın bir Pt tel) ve (3) Ag/AgCl referans elektrotu. Elektrot yüzeyinin yeterince Pt nanoyapısıyla kaplanması için elektrot potansiyeli 0,2 mM H2PtCl6 solüsyonunda -0.1 V'de yaklaşık 100 sn. boyunca sabit tutulmuştur.

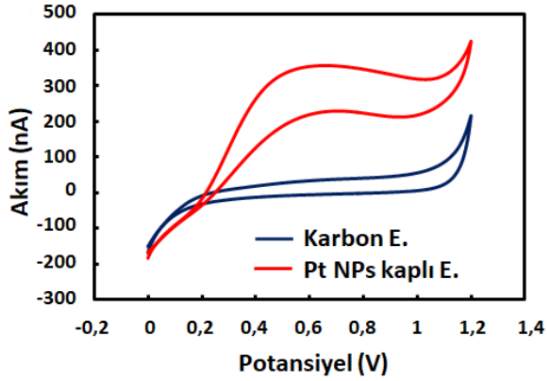

Şekil 2. KFE ve Pt-KFE'lerin 1mM H2O2 + PBS içerisinde elde edilen CV eğrileri

Elektrokimyasal kaplama işlemi sonucunda akım beklendiği üzere düşme eğilimi göstermiştir; akım 100

modifikasyonu elektrot yüzeyini arttırmış ve bu da akımda gözle görülür bir artışa neden olmuştur. Ayrıca, yüzeyin Pt ile başarılı bir şekilde modifiye edilip edilmediği bu elektrotların $O_2$'nin indirgenmesinde etkinliği araştırılarak ta ortaya konmuştur. Bu aşamada, Pt-KFE ve KFE'lerin PBS solüsyonu içerisinde +0,2 ve -0,4 V (vs. Ag/AgCl) aralığında CV eğrileri elde edilmiştir. Elde edilen sonuçlara göre iğne uçlu Pt-KFE $O_2$'nin indirgenmesinde KFE'e göre oldukça yüksek bir etkinlik göstermiştir. KFE ile -0,4 V'de (vs. Ag/AgCl) elde edilen akım seviyesi -19.8 nA iken bu seviye iğne uçlu Pt-KFE'ler ile -182 nA'e kadar çıkmıştır. Daha sonra, Pt-KFE'lerin SEM görüntüleri alınmış ve belli bir alan üzerinde EDS analizi de yapılmıştır. SEM görüntülerden ilki (Şekil 3A) elektrotun başarıyla üretildiğini göstermiş, ikinci SEM görüntüsü de yüzeyin elektrokimyasal Pt depozisyonu sonucunda çapı 50 nm'den düşük çeşitli boyutlarda NP ile kaplandığını göstermiştir (Şekil 3B). Yapılan enerji dağılım spektrometresi (EDS) analiz sonuçları da de bu nano yapıların % 91 oranında Pt olduğunu göstermektedir. Elde edilen sonuçlar KFE yüzeyinin Pt ile başarılı bir şekilde modifiye edildiğini kanıtlamaktadır (Şekil 3C). Daha sonra iğne uçlu Pt-KFE'lerin $H_2O_2$'nin algılanmasında gösterdiği etkinlik araştırılmıştır. İlk önce 1 mM $H_2O_2$ + PBS solüsyonunda iğne uçlu KFE ve Pt-KFE'lerin 0 ile 1,2 V (vs. Ag/AgCl) aralığında (50 mV/s hızda) CV eğrileri elde edilmiştir. (Şekil 4). İğne uçlu Pt-KFE'ler beklendiği KFE'lere göre H2O2'nin algılanmasında çok daha yüksek etkinlik göstermiştir.

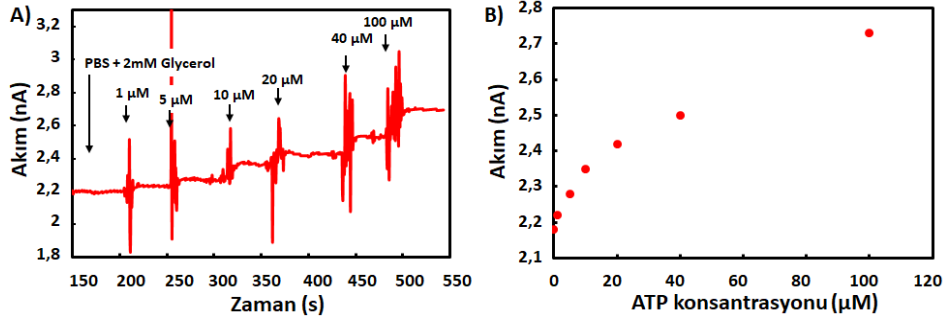

Şekil 5. Kronoamperometri ile farklı konsantrasyonda ATP algılanması (A) ve elde edilen sonuçlara göre çizilen kalibrasyon eğrisi (B).

sn'lik bir süre sonunda -400 nA seviyelerine kadar düşmüştür. İğne uçlu Pt-KFE'nin 1mM FMA içerisinde

Yaklaşık olarak +0.4 ile 1 V aralığında elde edilen akım $H_2O_2$ algılanması için en yüksek seviyeyi bulmuştur.

Dolayısıyla sonraki aşamalarda disk Pt-KFE'lerde olduğu gibi elektrokimyasal algılama için +0,7 V (vs. Ag/AgCl) kullanılmıştır. Bu voltajda iğne uçlu KFE ile elde edilen akım seviyesi 39 nA iken Pt-KFE'ler ile bu seviye 350 nA'ya kadar çıkmıştır. Bu değerler sadece daha kolay karşılaştırma için verilmiştir, gerçek değerlerin tepe akım noktası ile taban akım noktasının farkı şeklinde ifade edildiği göz önünde bulundurulmalıdır. En son aşamada disk-KFE tabanlı ATP biyosensörlerinde olduğu gibi iğne uçlu Pt-KFE yüzeyi gliserol kinaz ve gliserol fosfat oksidaz enzimleri ile modifiye edilerek biyosensörlerin üretimi gerçekleştirilmiştir (Şekil 1A). Bu aşamada, 5'er ünite (U) gliserol kinaz ve gliserol fosfat oksidaz enzimleri 0,1 M pirol + 0,7 KCl solüsyonuyla karıştırılmış ve böylece pirolün polipirole elektropolimerizasyonu şeklinde belirtilen enzimlerin elektrot yüzeyine immobilizasyonu yapılmıştır [13]. Yukarda da belirtildiği gibi elde edilen karışımın hacmi kullanılan yüksek enzim konsantrasyonuna bağlı olarak düşük olması (200 µL) nedeniyle elektropolimerizasyon bir Ag/AgCl teli ile 96'lık mikrotitrasyon plaklarının tek bir kuyucuğunda gerçekleştirilmiştir. Elektropolimerizasyon için CV kullanılmıştır; kısaca, uçları dH$_2$O'da yıkanmış elektrot probları referans elektrotu ile birlikte elektropolimerizasyon için hazırlanan enzim karışımının içine yerleştirilmiştir. Sonrasında elektrot potansiyeli 0 ile 0,8 V (vs. Ag/AgCl) arasında 50 mV/s hızında 8 kez taranarak iğne uçlu Pt-KFE'lerin yüzeyi belirtilen enzimlerle modifiye edilerek biyosensörlerin üretimi gerçekleştirilmiştir. Daha sonrasında üretilen biyosensörlerin ATP algılamasında gösterdiği yanıt elektrokimyasal olarak incelenmiştir. Algılama 2 mM gliserol içeren PBS solüsyonunda sabit +0,7 V (vs. Ag/AgCl) sabit voltajında kronoamperometri ile gerçekleştirilmiştir. İlk olarak disk Pt-KFE tabanlı ATP biyosensöründe olduğu gibi elde edilen sonucun doğruluğunu arttırmak için uygulanan sabit voltajda akımın doğrusal hale gelmesi beklenmiş ve sonrasında derişik ATP solüsyonu kullanılarak deteksiyon solüsyonunda bulunan ATP konsantrasyonu 0'dan 100 µM'a (sırasıyla 0, 5, 10, 20, 40 ve 100 µM) kadar sırasıyla adım adım artırılmıştır. Artan ATP konsantrasyonuna karşı akım değişimi zamana bağlı olarak kaydedilmiştir. Bu aşamada, polipirol ile immobilize edilen enzimlerin ATP ve gliserolü ardı sıra reaksiyonlarda kullanılması sonucu son ürün olarak elektrokimyasal ölçümü yapılabilen H$_2$O$_2$ üretmesi beklenmektedir (Şekil 1B). Elde edilen sonuçlara göre üretilen biyosensörün eklenen her bir ATP konsantrasyon seviyesine anlamlı cevap verdiği gözlemlenmiştir. Kronoamperometri grafiğine bakıldığında değişim rahatlıkla gözlemlenebilmektedir (Şekil 5A). Kalibrasyon eğrisinden de anlaşılacağı gibi her bir ATP konsantrasyon seviyesinde farklı bir akım değeri ölçülmüştür. Disk Pt-KFE tabanlı ATP biyosensöründen farklı olarak iğne uçlu Pt-KFE tabanlı ATP biyosensörü ile 1 µM ATP seviyesi anlamlı bir şekilde ölçülmüştür (S/N: 3) (Şekil 5B). Ayrıca, biyosensörün özellikle düşük konsantrasyonlarda (1-40 µM) yaklaşık olarak doğrusal cevap verdiği gözlemlenmiştir. ATP biyosensörünün pratikte kullanılabilmesi için fizyolojik olarak anlamlı asgari 1 – 50 µM konsantrasyon aralığında ATP'ye karşı anlamlı sonuç verebilmesi gerektiği düşünüldüğünde iğne uçlu Pt-KFE'lerin bu gibi çalışmalarda kullanımının uygun olduğu sonucuna varılabilir.

## 3. Bilgilendirme

## Kaynaklar

1. A. L. Buchachenko, D. A. Kuznetsov and N. N. Breslavskaya, Chemistry of Enzymatic ATP Synthesis: An Insight through the Isotope Window, Chemical Reviews, Vol. 112, p. 2042-2058, 2012.
2. B. Benziane, M. Björnholm, S. Pirkmajer, R. L. Austin, O. Kotova, B. Viollet, J. R. Zierath and A. V. Chibalin, Activation of AMP-activated Protein Kinase Stimulates Na(+),K(+)-ATPase Activity in Skeletal Muscle Cells, The Journal of Biological Chemistry, Vol. 287, p. 23451-23463, 2012.
3. M. F. Jarvis and B. S. Khakh, ATP-gated P2X cation-channels, Neuropharmacology, Vol. 56, p. 208-215, 2009.
4. M. Şen, K. Ino, K. Y. Inoue, T. Arai, T. Nishijo, A. Suda, R. Kunikata, H. Shiku and T. Matsue, LSI-based amperometric sensor for real-time monitoring of embryoid bodies, Biosensors and Bioelectronics, Vol. 48, p. 12-18, 2013.
5. K. Furuya, M. Sokabe and R. Grygorczyk, Real-time luminescence imaging of cellular ATP release, Methods, Vol. 66, p. 330-344, 2014.
6. E. Baldrich, A. Restrepo and C. K. O'Sullivan, Aptasensor Development: Elucidation of Critical Parameters for Optimal Aptamer Performance, Analytical Chemistry, Vol. 76, p. 7053-7063, 2004.
7. P. Hong, W. Li and J. Li, Applications of Aptasensors in Clinical Diagnostics, Sensors, Vol. 12, p. 1181, 2012.
8. M. Şen, K. Ino, H. Shiku and T. Matsue, A new electrochemical assay method for gene expression using hela cells with a secreted alkaline phosphatase (SEAP) reporter system, Biotechnology and Bioengineering, Vol. 109, p. 2163-2167, 2012.
9. M. Şen and A. Demirci, pH-Dependent ionic-current-rectification in nanopipettes modified with glutaraldehyde cross-linked protein membranes, RSC Advances, Vol. 6, p. 86334-86339, 2016.
10. M. Sen, Fabrication of Ultamicro Carbon Fiber Electrode Probes for Detection of O2 and H2O2 Istanbul University - Journal of Electrical & Electronics Engineering, Vol. 17, p. 3049-3055, 2017.
11. A. Bayram, M. Serhatlioglu, B. Ortac, S. Demic, C. Elbuken, M. Sen and M. E. Solmaz, Integration of glass micropipettes with a 3D printed aligner for microfluidic flow cytometer, Sensors and Actuators A: Physical, Vol. 269, p. 382-387, 2018.

12. M. Şen, Y. Takahashi, Y. Matsumae, Y. Horiguchi, A. Kumatani, K. Ino, H. Shiku and T. Matsue, Improving the Electrochemical Imaging Sensitivity of Scanning Electrochemical Microscopy-Scanning Ion Conductance Microscopy by Using Electrochemical Pt Deposition, Analytical Chemistry, Vol. 87, p. 3484-3489, 2015.
13. V. K. Aydin and M. Şen, A facile method for fabricating carbon fiber-based gold ultramicroelectrodes with different shapes using flame etching and electrochemical deposition, Journal of Electroanalytical Chemistry, Vol. 799, p. 525-530, 2017.